\documentclass{article}
\usepackage{waspaa21,amsmath,graphicx,url,times}
\usepackage{color}
\usepackage{amssymb} 
\usepackage{multirow}
\usepackage{float} 
\usepackage[position=bottom]{subfig} 
\usepackage[group-separator={,}]{siunitx} 

\title{Disentanglement Learning for Variational Autoencoders \\ Applied to Audio-Visual Speech Enhancement}

\name{Guillaume Carbajal,\thanks{This work has been funded by the German Research Foundation (DFG) in the transregio project Crossmodal Learning (TRR 169) and ahoi.digital.}
      Julius Richter,
      Timo Gerkmann}
\address{Signal Processing (SP), Universität Hamburg, Germany \\
\{guillaume.carbajal, julius.richter, timo.gerkmann\}@uni-hamburg.de}

\begin{document}

\ninept
\maketitle

\begin{sloppy}

\begin{abstract}
    Recently, the standard variational autoencoder has been successfully used to learn a probabilistic prior over speech signals, which is then used to perform speech enhancement. Variational autoencoders have then been conditioned on a label describing a high-level speech attribute (e.g. speech activity) that allows for a more explicit control of
    speech generation.
    However, the label is not guaranteed to be disentangled from the other latent variables, which results in limited performance improvements compared to the standard variational autoencoder. In this work, we propose to use an adversarial training scheme for variational autoencoders to disentangle the label from the other latent variables. At training, we use a discriminator that competes with the encoder of the variational autoencoder. Simultaneously, we also use an additional encoder that estimates the label for the decoder of the variational autoencoder, which proves to be crucial to learn disentanglement.
    We show the benefit of the proposed disentanglement learning when a voice activity label, estimated from visual data, is used for speech enhancement.
\end{abstract}


\section{Introduction}
\label{sec:intro}

Single-channel speech enhancement consists in recovering a speech signal from a mixture signal captured with one microphone in a noisy environment \cite{vincent_audio_2018}. Common speech enhancement approaches estimate the speech signal using a filter in the time-frequency domain to reduce noise while avoiding speech artifacts \cite{hendriks_dft-domain_2013}. Under the Gaussian assumption, the optimal filter in the minimum mean square error sense requires estimating the signal variances \cite{breithaupt_cepstral_2007, fevotte_nonnegative_2009, gerkmann_unbiased_2012}.

Recently, deep generative models based on the variational autoencoder (VAE)  have gained attention for learning the probability distribution of complex data \cite{kingma_introduction_2019}. VAEs have been used to learn a prior distribution of clean speech, and have been combined with an untrained non-negative matrix factorization (NMF) noise model to estimate the signal variances using a Monte Carlo expectation maximization (MCEM) algorithm \cite{bando_statistical_2018, leglaive_variance_2018}. 
However, since the VAE speech model is trained with clean speech only in an unsupervised manner, the speech prior generates a speech-like signal when only noise is present.
As a result, 
the VAE often outputs speech-like noise when applied to noisy speech \cite{bando2020adaptive}. 
%

An increased robustness can be obtained by incorporating temporal dependencies \cite{richterSpeechEnhancementStochastic2020a}, noise-aware training \cite{fang2021noiseaware}, or by conditioning the VAE on a label describing an attribute of the data that allows for a more explicit control of data generation \cite{kingma_semi-supervised_2014}. For various speech-related tasks, VAEs have been conditioned on a label describing a speech attribute, such as speaker identity \cite{kameoka_supervised_2019, du_semi-supervised_2021}, phoneme \cite{du_semi-supervised_2021} or speech activity \cite{carbajal_guided_2021}. Ideally, the label should be independent from the other latent dimensions to obtain an explicit control of speech generation. However, the semi-supervised learning of conditional VAEs does not guarantee that it will promote independence between the label and the other latent dimensions \cite{siddharth_learning_2017}. As a result, the speech generation can only partially be controlled by the label.



Disentanglement learning aims at making all the dimensions of the latent distribution independent from each other \cite{higgins_beta_2017, kim_disentangling_2018, chen_isolating_2018}, e.g. by holding different axes of variation fixed during training \cite{mathieu_disentangling_2019}.  Semi-supervised disentanglement approaches, on the other hand, tackle the problem of making only some observed (often interpretable) variations in the data independent from the other latent dimensions which themselves remain entangled \cite{kingma_semi-supervised_2014, siddharth_learning_2017, locatello_disentangling_2020}. In particular, some of them rely on adversarial training in the latent space for image generation \cite{lample2017fader, creswell_adversarial_2018}.

In this work, we propose to use a semi-supervised disentanglement approach based on adversarial training in the latent space. At training, we use a discriminator that competes with the encoder of the VAE. The discriminator aims at identifying the label from the other latent dimensions, whereas the encoder aims at making it unable to estimate the label. Simultaneously, we also use an additional encoder that estimates the label for the decoder of the VAE, which proves to be crucial to learn disentanglement during training.
We show the benefit of the proposed disentanglement learning when a voice activity label, estimated from visual data, is used for speech enhancement.


The rest of this paper is organized as follows.  In Section \ref{sec:background} we summarize the background related to the VAE for speech enhancement. Section \ref{sec:proposed} describes our proposed approach. The experimental setup is described in Section \ref{sec:experimental} which is followed by the evaluation in Section \ref{sec:results}.

\section{Background}
\label{sec:background}

\subsection{Mixture model and filtering}

In the time-frequency domain using the short time Fourier transform (STFT), the mixture signal $x_{nf} \in \mathbb{C}$ is the sum of the clean speech $s_{nf} \in \mathbb{C}$ and the noise $b_{nf} \in \mathbb{C}$:
\begin{align}
x_{nf} = \sqrt{g_n}\, s_{nf} + b_{nf},
\label{eq:mixture_model}
\end{align}
at time frame index $n \in [1, N]$ and frequency bin $f \in [1,F]$, where $N$ denotes the number of time frames and $F$ the number of frequency bins of the utterance. The scalar  $g_n \in \mathbb R_{+}$ represents a frequency-independent but time-varying gain providing some robustness with respect to the time-varying loudness of different speech signals \cite{leglaive_variance_2018}.

Under the Gaussian assumption, the clean speech $s_{nf}$ can be estimated in the minimum mean square error sense using the Wiener estimator:
\begin{equation}
\widehat{s}_{nf} = \frac{\widehat{g}_n \widehat{v}_{s, nf}}{\widehat{g}_n \widehat{v}_{s, nf} + \widehat{v}_{b, nf}} \, x_{nf},
\label{eq:wiener}
\end{equation}
where $\widehat{v}_{s, nf}$ and $\widehat{v}_{b,nf}$ are the estimated variances of the clean speech $s_{nf}$ and the noise $b_{nf}$, respectively. Under a local stationary assumption, short-time power spectra $|s_{nf}|^2$ and $|b_{nf}|^2$ are unbiased estimates of the signal variances \cite{liutkus_gaussian_2011}.

\subsection{VAE as speech prior}
\label{sec:vae}

\subsubsection{Standard VAE}

\begin{figure}[b]
\centering
\begin{minipage}[b]{.8\columnwidth}
  \centering
  \centerline{\includegraphics[width=\textwidth]{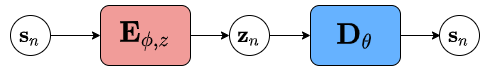}}
  \centerline{(a) Standard VAE}\medskip
\end{minipage}
\\
\begin{minipage}[b]{.8\columnwidth}
  \centering
  \centerline{\includegraphics[width=\textwidth]{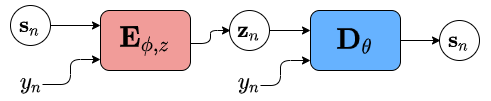}}
  \centerline{(b) Conditional VAE \cite{carbajal_guided_2021}}\medskip
\end{minipage}
\caption{Training scheme for (a) the standard VAE and (b) the conditional VAE. \label{fig:model_M1}}
\end{figure}

 
The standard VAE is used to learn a prior over clean speech \cite{bando_statistical_2018, leglaive_variance_2018}. Fig.~\ref{fig:model_M1}a shows the training scheme of the  standard VAE. At time frame $n$, the frequency bins of clean speech $\mathbf s_{n} \in \mathbb C^F$ are modeled as
\begin{equation}
    p_\theta(\mathbf{s}_n | \mathbf{z}_{n}) = \mathcal{CN}(\mathbf{0}, \text{diag}(\mathbf{D}_{\theta}(\mathbf{z}_{n}))), 
    \quad \mathbf{z}_{n} \sim \mathcal{N}(\mathbf 0, \mathbf I),
\end{equation}
where $\mathbf z_{n} \in \mathbb R^L$ denotes a latent variable of dimension $L$ and $\mathbf{D}_{\theta}(\mathbf{z}_n) \in \mathbb{R}_+^F$  is the output of the \emph{decoder} network $\mathbf{D}_{\theta}(\cdot)$ pa\-ram\-e\-trized by $\theta$.

In variational inference, the posterior of $\mathbf{z}_{n}$ is approximated as
\begin{equation}
    q_\phi(\mathbf{z}_{n} |\ \mathbf{s}_n) = \mathcal N( \boldsymbol{\mu}_{\phi}(|\mathbf{s}_n|^2), \operatorname{diag}( \mathbf{v}_{\phi}(|\mathbf{s}_n|^2))) \label{eq:encoder_m1},
\end{equation}
where $\boldsymbol \mu_\phi(|\mathbf{s}_n|^2) \in \mathbb R^D$ and $\mathbf v_\phi(|\mathbf{s}_n|^2) \in \mathbb R_+^D$ are the outputs of the \emph{encoder} network $\mathbf{E}_{\phi,z}(\cdot)$ parametrized by $\phi$.

The encoder $\mathbf{E}_{\phi,z}(\cdot)$ and decoder $\mathbf{D_{\theta}}(\cdot)$ are jointly trained by maximizing a lower bound of the marginal per-frame log-like\-lihood, called the evidence lower bound (ELBO):
\begin{equation}
      \mathcal{L}_\text{ELBO} = \mathbb E_{q_\phi(\mathbf{z}_{n}| \mathbf{s}_n)} [\log p_\theta (\mathbf{s}_n| \mathbf{z}_{n})] -  \mathcal{D}_\text{KL}(q_\phi (\mathbf{z}_{n}| \mathbf{s}_n)|| p(\mathbf{z}_{n})),
\end{equation}
where the first term is the reconstruction loss and $ \mathcal{D}_\text{KL}(\cdot||\cdot)$ denotes the Kullback-Leibler divergence. However, since the training loss $\mathcal{L}_\text{ELBO}$ is unsupervised, there is no explicit control of speech generation.

\subsubsection{Conditional VAE}

The VAE can be conditioned on a label $y_n \in \mathcal{Y}$ describing a speech attribute (e.g. speech activity) that allows for a more explicit control of speech generation  \cite{carbajal_guided_2021}. A common approach is to make use of the label $y_n$ by directly inputting it in both the encoder $\mathbf{E}_{\phi,z}(|\mathbf{s}_n|^2, y_n)$ and the decoder $\mathbf{D}_{\theta}(\mathbf{z}_n, y_n)$ (see Fig \ref{fig:model_M1}b) \cite{kameoka_supervised_2019, du_semi-supervised_2021, carbajal_guided_2021}. The training loss remains the same as for the standard VAE, i.e. $\mathcal{L_\text{ELBO}}$.

\subsection{Non-negative matrix factorization as noise model}

As in Leglaive et al. \cite{leglaive_variance_2018}, the noise variance is modeled with an untrained NMF as 
\begin{equation}
    v_{b,nf} = \left\{\mathbf H \mathbf W \right\}_{nf},
\end{equation}
where $\mathbf H \in \mathbb R_+^{N \times K}$ and $\mathbf W \in \mathbb R_+^{K \times F}$ are two non-negative matrices representing the temporal activations and spectral patterns of the noise power spectrogram. $K$ denotes the NMF rank.

\subsection{Clean speech estimation}
\label{sec:speech_est}
\begin{figure}[ht]
\centering
\includegraphics[width=0.8\columnwidth]{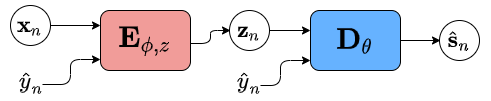}

\caption{Estimation of the speech variance $v_{s,nf}$ at test time.}
\label{fig:test_time}
\end{figure}
Speech variance estimation at test time is shown in Fig. \ref{fig:test_time}. First, a pretrained noise-robust classifier which is trained separately from the VAE provides the label estimate $\widehat{y}_{n}$ from the mixture $\mathbf{x}_{n}$ as input for the VAE. The outputs of the encoder and decoder are then $\mathbf{E}_{\phi,z}(|\mathbf{x}_n|^2, \widehat{y}_n)$ and $\mathbf{D}_{\theta}(\mathbf{z}_n, \widehat{y}_n)$, respectively. Given the speech prior provided by the VAE and the noise model, the mixture signal $x_{nf}$ is distributed as
\begin{equation}
x_{nf} |\ \mathbf{z}_{n} \sim \mathcal{CN}(0, {g}_n \{\mathbf D_{\theta} (\mathbf{z}_{n}, \widehat{y}_n)\}_f + \{\mathbf H \mathbf W\}_{nf})
\label{eq:likelihood_mixture}
\end{equation}
where $\Theta_u = \{{g}_n, {\mathbf H}, {\mathbf W} \}$ are the unsupervised parameters to be estimated. Since the resulting optimization problem is intractable due to the non-linear relation between the speech variance $v_{s,nf}$ and the latent variable $\mathbf{z}_n$, an MCEM algorithm is employed to iteratively optimize the unsupervised parameters $\Theta_u$ \cite{leglaive_variance_2018}. At each iteration, the estimated terms $\widehat{g}_n \{\mathbf D_{\theta} (\mathbf{z}_{n})\}_f$ and $\{\widehat{\mathbf H} \widehat{\mathbf W}\}_{nf}$ are supposed to get closer to the true variances $v_{s,nf}$ and $v_{b,nf}$, respectively.

Ideally, in order to obtain an explicit control of speech generation, the label $y_n$ should be independent from the latent variable $\mathbf{z}_n$. However, the training loss $\mathcal{L}_\text{ELBO}$ does not guarantee that it will promote independence between the label $y_n$ and the latent variable $\mathbf{z}_n$ \cite{siddharth_learning_2017}. As a result, speech generation can only partially be controlled by the label $y_n$.

\section{Disentanglement learning}
\label{sec:proposed}

Inspired by recent work on semi-supervised disentanglement learning \cite{lample2017fader, creswell_adversarial_2018}, we propose to use an adversarial training scheme on the latent space to disentangle the label $y_n$ from the latent variable $\mathbf{z}_n$.  More particularly, we propose a different training loss for the encoder of the VAE.
\subsection{Architecture}

\begin{figure}[ht]
\centering
\includegraphics[width=0.8\columnwidth]{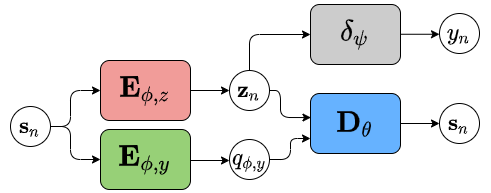}

\caption{Adversarial training scheme for the conditional VAE }
\label{fig:adv_cvae}
\end{figure}

Fig. \ref{fig:adv_cvae} shows the adversarial training scheme. At training, we use a discriminator $\boldsymbol{\delta}_\psi(\cdot)$ that competes with the encoder $\mathbf{E}_{\phi,z}(\cdot)$ of the VAE, which we refer to as the \emph{adversarial-encoder}. The discriminator $\boldsymbol{\delta}_\psi(\cdot)$, parametrized by $\psi$, aims at identifying the label $y_n$ from the latent variable $\mathbf{z}_n$ whereas the \emph{adversarial-encoder} $\mathbf{E}_{\phi,z}(\cdot)$ aims at making it unable to estimate the label $y_n$. Simultaneously, we also estimate the label $y_n$ using a classifier that we denote as the \emph{classifier-encoder} $\mathbf{E}_{\phi,y}(\cdot)$, parametrized by $\phi$.

\textbf{Label $y_n$}\; We consider voice activity detection (VAD) for the label, i.e. $y_n \in \{0, 1\}$, which is here obtained from visual data.

\textbf{Discriminator $\boldsymbol \delta_\psi(\cdot)$} \;
Similarly to Fader Networks for image generation \cite{lample2017fader}, the discriminator $\boldsymbol \delta_\psi(\cdot)$ estimates the probability that the label $y_n=1$, i.e. $p_\psi:= P(y_n=1|\mathbf{z}_n) = \boldsymbol \delta_\psi(\mathbf z_n)$. We use the binary cross entropy (BCE) as the learning objective:
\begin{equation}
    \mathcal{L}_\text{dis} = y_n \log p_\psi + (1 - y_n) \log (1 - p_\psi).
\end{equation}

\textbf{Adversarial-Encoder $\mathbf E_{\phi,z}(\cdot)$}\; To force the discriminator $\boldsymbol \delta_\psi(\cdot)$ to make incorrect classification, adversarial approaches in image generation proposed to use $\mathcal{L}_\text{adv-enc} = - \mathcal{L}_\text{dis}$ as the training loss for the \emph{adversarial-encoder} $\mathbf E_{\phi,z}(\cdot)$ \cite{lample2017fader, creswell_adversarial_2018}. However, this loss does not make the discriminator $\boldsymbol \delta_\psi(\cdot)$ unable to predict the label $y_n$. Instead, we thus propose to minimize the negative binary entropy function with $q_{\phi, z} := P(y_n = 1| \mathbf s_s) = \boldsymbol \delta_\psi(\mathbf{E}_{\phi,z}(|\mathbf{s}_n|^2))$:
\begin{equation}
    \begin{split}
    \mathcal{L}_\text{adv-enc} &= q_{\phi, z} \log q_{\phi, z} + (1 - q_{\phi, z}) \log (1- q_{\phi, z}).
    \end{split}
\end{equation}

\textbf{Classifier-Encoder $\mathbf E_{\phi,y}(\cdot)$} \; Similarly to Creswell et al. for image generation \cite{creswell_adversarial_2018}, we estimate the label $y_n$ for the decoder $\mathbf{D}_\theta(\cdot)$ using the \emph{classifier-encoder} $\mathbf E_{\phi,y}(\cdot)$ as $q_{\phi,y} := P(y_n=1|\mathbf s_n) = \mathbf E_{\phi,y}(|\mathbf{s}_n|^2)$. We use the BCE as the learning objective:
\begin{equation}
    \mathcal{L}_\text{clf-enc} = y_n \log q_{\phi,y} + (1 - y_n) \log (1 - q_{\phi,y}).
\end{equation}

\textbf{Decoder $\mathbf D_{\theta}(\cdot)$}\; Instead of using the binary label $y_n$ as an input to the decoder such that $\mathbf{D}_\theta(\mathbf{z}_n, y_n)$, we use the posterior probability $q_{\phi,y}$ as a soft value, such that $\mathbf D_{\theta}(\mathbf{z}_n, q_{\phi,y})$.

\subsection{Adversarial training}

The complete loss of the conditional VAE is:
\begin{equation}
    \mathcal{L}_\text{adv-VAE} = \mathcal{L}_\text{ELBO} + \alpha \mathcal{L}_\text{adv-enc} + \beta \mathcal{L}_\text{clf-enc}.
\end{equation}
At each step of the training, we update the networks as follows:
\begin{enumerate}
    \item Update $\mathbf E_{\phi,z}(\cdot)$, $\mathbf E_{\phi,y}(\cdot)$ and $\mathbf D_{\theta}(\cdot)$ using $\mathcal{L}_\text{adv-VAE}$
    \item Update $\boldsymbol \delta_\psi(\cdot)$ using $\alpha \mathcal{L}_\text{dis}$\footnote{$\alpha$ is related to $\mathcal{L}_\text{adv-enc}$ and proved to provide better results in practice.}
\end{enumerate}

\subsection{Clean speech estimation}
\begin{figure}[ht]
\centering
\includegraphics[width=0.8\columnwidth]{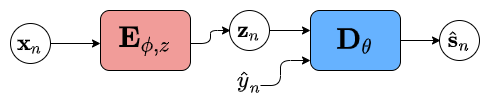}
\caption{Estimation of the speech variance $v_{s,nf}$ at test time.}
\label{fig:adv_test_time}
\end{figure}

Speech variance estimation at test time is shown in Fig. \ref{fig:adv_test_time}. At test time, we estimate the label $y_n$ with a pretrained noise-robust classifier which is trained separately from the VAE on visual data. For estimating the unsupervised parameters $\Theta_u$, we use the same MCEM configuration as in Section \ref{sec:speech_est}.


\section{Experimental setup}
\label{sec:experimental}

\hspace{\parindent} \textbf{Dataset} \; For training, validation, and test we used the NTCD-TIMIT dataset \cite{abdelaziz_ntcd-timit_2017}, which consists of audio-visual recordings. The visual data consist of videos of the lip region at $30$~frames-per-second (FPS). All audio signals have a sampling rate of $16$~\text{kHz}. In addition to clean speech signals, we use the noisy versions, with 2 types of stationary noise \{\textit{Car}, \textit{White}\} and four types of nonstationary noises \{\textit{Living Room}, \textit{Cafe}, \textit{Babble}, \textit{Street}\}. We evaluate on 3 signal-to-noise ratios (SNRs): \{$0$, $+5$, $+10$\}~dB. Note that these levels were computed on speech segments only.
For training, the clean-speech and visual sets
used are of about $5$~h. For the test, the noisy speech subset consists of \num{15876} utterances of about $5$~s each.

\textbf{Baselines} \; For the noise-robust classifier estimating the label $y_n$ at test time, 
we use the visual-only variant of Ariav and Cohen's audio-visual VAD classifier \cite{ariav_end-to-end_2019}. We pretrain the classifier on the visual training set of the NTCD-TIMIT dataset.
For the baselines, we use the standard and the conditional VAEs, which we denote as \emph{VAE} and \emph{CVAE}, respectively, both trained using $\mathcal{L}_\text{ELBO}$.

\textbf{Hyperparameter settings} \;  The STFT is computed using a $64\, \text{ms}$ Hann window with 75\% overlap, resulting in $F=513$ unique frequency bins and an audio frame rate of $62.5$ FPS. To obtain the same visual frame rate as the audio frame rate, we use FFMPEG \cite{ffmpeg}. This guarantees that the audio and visual frames are synchronized. To obtain the ground truth for the VAD label $y_n$, we use simple thresholding in the time domain.

For a fair comparison between all the approaches, we consider a similar architecture for each subnetwork. Tab. \ref{tab:model_configurations} shows the configuration of the models. We set the dimension of the latent space to $L=16$. For the loss, we empirically found that $\alpha=\beta=10$ provides excellent performance. We use the Adam optimizer with standard configuration \cite{kingma_adam_2014}.  We set the batch size to $128$. Early stopping with a patience of $10$ epochs is performed using $\mathcal{L}_\text{adv-enc}$ on the validation set. For the MCEM we follow the settings of Leglaive et al. and set the NMF rank to $K=10$ \cite{leglaive_variance_2018}.

\begin{table}[b]
\centering
\begin{tabular}{l | c c c | c}
& \multicolumn{3}{c|}{Hidden layers} & Output layer \\
\textbf{Subnetwork} & \textbf{\# layers} & \textbf{\# units} & \textbf{act. fn} & \textbf{act. fn} \\
\hline
$\boldsymbol \delta_{\psi}(\cdot)$ & $2$ & $128$ & $\operatorname{ReLU}$ & $\operatorname{sigmoid}$ \\
$\mathbf{E}_{\phi,z}(\cdot)$ & $2$ & $128$ & $\operatorname{tanh}$ & $\operatorname{identity}$ \\
$\mathbf{E}_{\phi,y}(\cdot)$ & $2$ & $128$ & $\operatorname{ReLU}$ & $\operatorname{sigmoid}$ \\
$\mathbf{D}_{\theta}(\cdot)$ & $2$ & $128$ & $\operatorname{tanh}$ & $\operatorname{exp}$
\end{tabular} 
\caption{Model configurations.}
\label{tab:model_configurations}
\end{table}

\textbf{Metrics} \; To evaluate speech enhancement  performance, we use the scale-invariant signal-to-distortion ratio (SI-SDR) measured in dB \cite{roux_sdr_2019}, raw scores of the extended short-time objective intelligibility (ESTOI) with values between $0$ and $1$ \cite{jensen_algorithm_2016}, and
the perceptual objective listening quality analysis (POLQA) score with values between $1$ and $5$ \cite{polqa}.
To evaluate the classification performance of the visual-only VAD classifier used at test time, we use the F1-score which combines the precision and recall rates.

\begin{table}[t]
\centering
\setlength{\tabcolsep}{4pt}
\begin{tabular}{l | c c c}
\textbf{Model}& \textbf{SI-SDR} (dB) & \textbf{ESTOI} & \textbf{POLQA} \\
 \hline
Mixture & $-1.3 \pm 0.1$ & $0.38 \pm 0.00$ & $1.40 \pm 0.01$\\  
 \hline
\emph{VAE} + $\mathcal{L}_\text{ELBO}$& $\;\;\;3.9 \pm 0.1$ & $0.36 \pm 0.00$ & $1.52 \pm 0.01$\\  
\hline
\emph{CVAE} + $\mathcal{L}_\text{ELBO}$ &$\;\;\ 4.6 \pm 0.1$ & $0.38 \pm 0.00  $ & $\mathbf{1.57 \pm 0.01}$ \\  
\hline
\emph{CVAE} + $\mathcal{L}_\text{adv-VAE}$ & $\;\;\ \mathbf{5.3 \pm 0.1}$ & $\mathbf{0.38 \pm 0.00}$ & $1.56 \pm 0.01$ \\
\hline
\end{tabular}
\caption{Average performance for nonstationary noises. ESTOI and POLQA are only evaluated during speech activity while the proposed disentanglement yields strong improvements in speech absence.}
\label{tab:results_nonstationary}
\end{table}

\begin{table}[t]
\centering
\setlength{\tabcolsep}{4pt}
\begin{tabular}{l | c c c}
\textbf{Model}& \textbf{SI-SDR} (dB) & \textbf{ESTOI} & \textbf{POLQA} \\
 \hline
Mixture & $-6.8 \pm 0.2$ & $0.46 \pm 0.00$ & $1.65 \pm 0.02$\\
 \hline
\emph{VAE} + $\mathcal{L}_\text{ELBO}$ &\;\;\  $6.0 \pm 0.2$ & $0.42 \pm 0.00$ & $1.57 \pm 0.01$\\  
\hline
\emph{CVAE} + $\mathcal{L}_\text{ELBO}$ &\;\;\  $ \mathbf{7.7 \pm 0.1}$ & $\mathbf{0.46 \pm 0.00}  $ & $\mathbf{1.67 \pm 0.01}$ \\  
\hline
\emph{CVAE} + $\mathcal{L}_\text{adv-VAE}$ &\;\;\ $6.8 \pm 0.1$ & $0.43 \pm 0.00$ & $1.57 \pm 0.01$ \\
\hline
\end{tabular}
\caption{Average performance for stationary noises.}
\label{tab:results_stationary}
\end{table}

\section{Results}
\label{sec:results}


\subsection{Average performance}

The visual-only VAD classifier gives F1-score = 88\% on the test set. Fig.~\ref{fig:spectrograms} shows an example of the reconstructed spectrograms by the VAEs, i.e. without using the MCEM algorithm. The proposed \emph{CVAE} + $\mathcal{L}_\text{adv-VAE}$ is the only approach with a meaningful output when speech absence is detected, i.e. $\widehat{y}_n=0$. This illustrates that disentanglement works with the proposed approach.

Tab. \ref{tab:results_nonstationary} shows the average results for nonstationary noises. \emph{CVAE} + $\mathcal{L}_\text{adv-VAE}$ outperforms both \emph{VAE} + $\mathcal{L}_\text{ELBO}$ and \emph{CVAE} + $\mathcal{L}_\text{ELBO}$ in terms of SI-SDR. This is again explained by our successful disentanglement which prevents \emph{CVAE} + $\mathcal{L}_\text{adv-VAE}$ from outputting a noise-like signal when speech absence is detected, i.e. $\widehat{y}_n=0$. Conversely, \emph{VAE} + $\mathcal{L}_\text{ELBO}$ and \emph{CVAE} + $\mathcal{L}_\text{ELBO}$ often outputs speech-like noise when $\widehat{y}_n=0$. Note that, unlike SI-SDR, the benefits of the proposed approach can not be visible
with ESTOI and POLQA because these metrics are only computed in speech presence.

Tab. \ref{tab:results_stationary} shows the average results for stationary noises. \emph{CVAE} + $\mathcal{L}_\text{adv-VAE}$ outperforms \emph{VAE} + $\mathcal{L}_\text{ELBO}$ in terms of SI-SDR. However, \emph{CVAE} + $\mathcal{L}_\text{adv-VAE}$ is outperfomed by \emph{CVAE} + $\mathcal{L}_\text{ELBO}$. This is also explained by our successful disentanglement which forces \emph{CVAE} + $\mathcal{L}_\text{adv-VAE}$ to output a signal when speech presence is detected, i.e. $\widehat{y}_n=1$. Since stationary noise is present on the entire utterance, unlike \emph{VAE} + $\mathcal{L}_\text{ELBO}$ and \emph{CVAE} + $\mathcal{L}_\text{ELBO}$, \emph{CVAE} + $\mathcal{L}_\text{adv-VAE}$ is forced to output both speech and noise when speech presence is detected which is not always beneficial, but can likely be addressed by more informative labels in future work. Code and audio examples are available online\footnote{\url{https://uhh.de/inf-sp-disentangled2021}}.




\subsection{Analysis of the training scheme}

While the above results show the performance of \emph{CVAE} with the proposed training scheme using $\mathcal{L}_\text{adv-VAE}$, we need further analysis regarding what contributes to the performance of the proposed training scheme \emph{CVAE} + $\mathcal{L}_\text{adv-VAE}$. Since the proposed disentanglement yields strong improvements in speech absence while ESTOI and POLQA are only evaluated during speech presence, we only consider SI-SDR here.

Tab. \ref{tab:training_scheme} shows the results on average and per noise stationarity with variants of the proposed training scheme. The first line corresponds to the proposed training scheme using \emph{CVAE} + $\mathcal{L}_\text{adv-VAE}$. The second and third lines correspond to the case where the binary label $y_n$ is used as an input to the decoder $\mathbf{D}_\theta(\mathbf{z}_n, y_n)$, instead of the soft value $q_{\phi,y}$. These training scheme variants are outperformed by the initially-proposed training scheme \emph{CVAE} + $\mathcal{L}_\text{adv-VAE}$ in terms of SI-SDR. We can also observe this on Fig.\ \ref{fig:cvae_ldis}. Thus, we conclude that estimating $q_{\phi,y}$ using the \emph{classifier-encoder} $\mathbf{E}_{\phi,y}(\cdot)$ as an input to the decoder $\mathbf{D}_\theta(\mathbf{z}_n, q_{\phi,y})$ is crucial to learn disentanglement.





\begin{figure}[t]
\vspace{-.5cm}
\centering
\subfloat[Mixture\label{fig:mixture}]{
\centering
\includegraphics[width=.315\columnwidth]{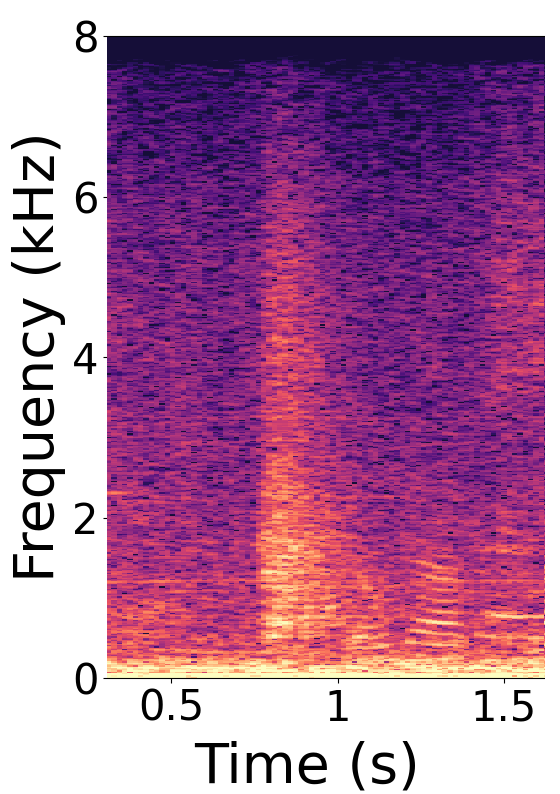}
}
\subfloat[Clean speech\label{fig:clean}]{
\centering
\includegraphics[width=.251\columnwidth]{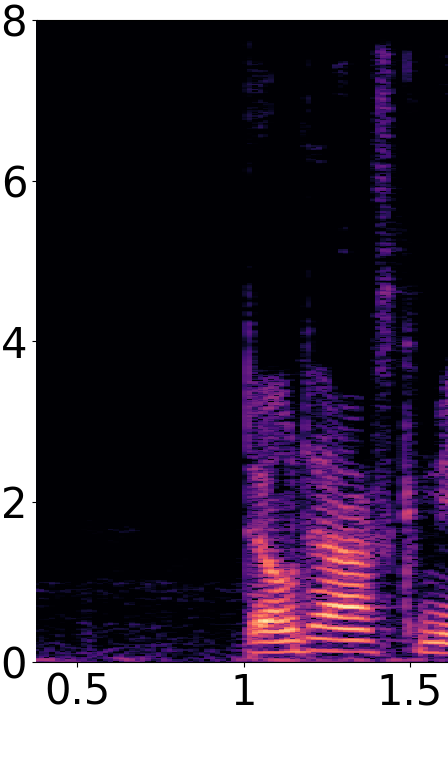} 
}
\subfloat[\emph{VAE} + $\mathcal{L}_\text{ELBO}$ \label{fig:recon_stcn}]{
\centering
\includegraphics[width=.405\columnwidth]{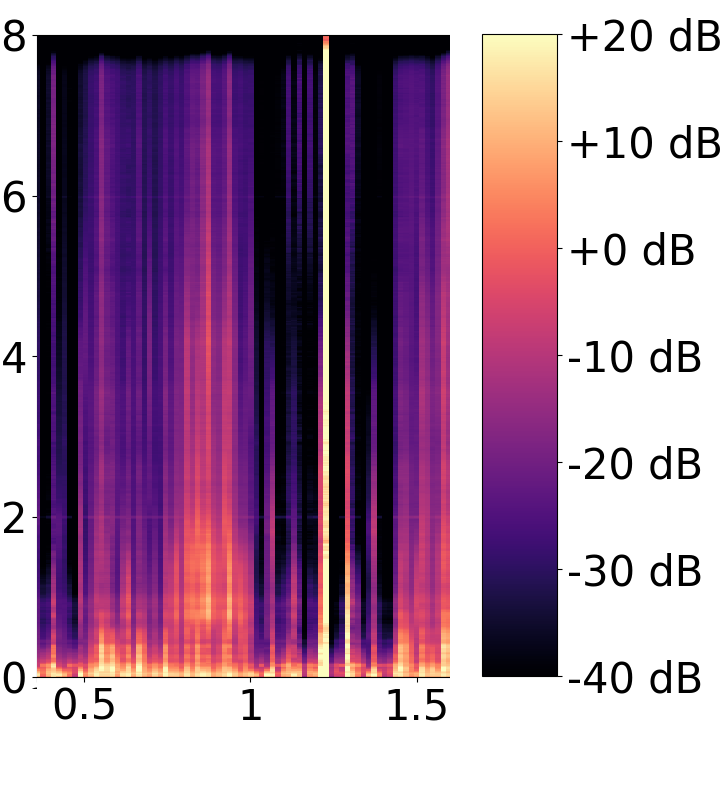}
}
\\
\subfloat[\emph{CVAE} + $\mathcal{L}_\text{ELBO}$\label{fig:cvae_elbo}]{
\centering
\includegraphics[width=.248\columnwidth]{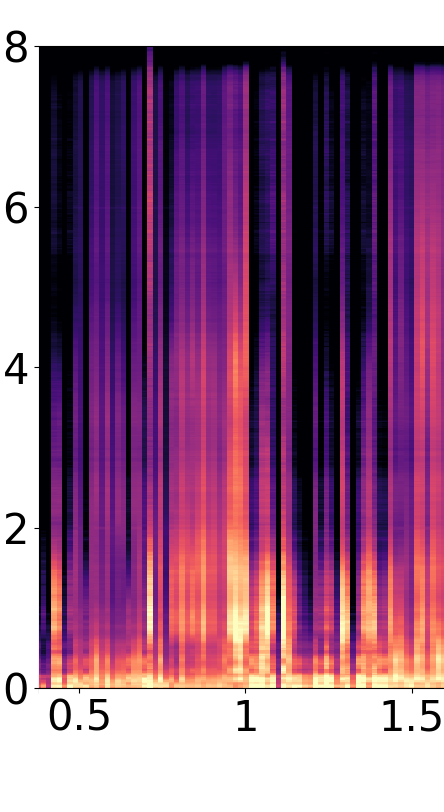}
}
\hspace{.22cm}
\subfloat[\emph{CVAE} + $\mathcal{L}_\text{adv-VAE}$ \label{fig:speech2}]{
\centering
\includegraphics[width=.272\columnwidth]{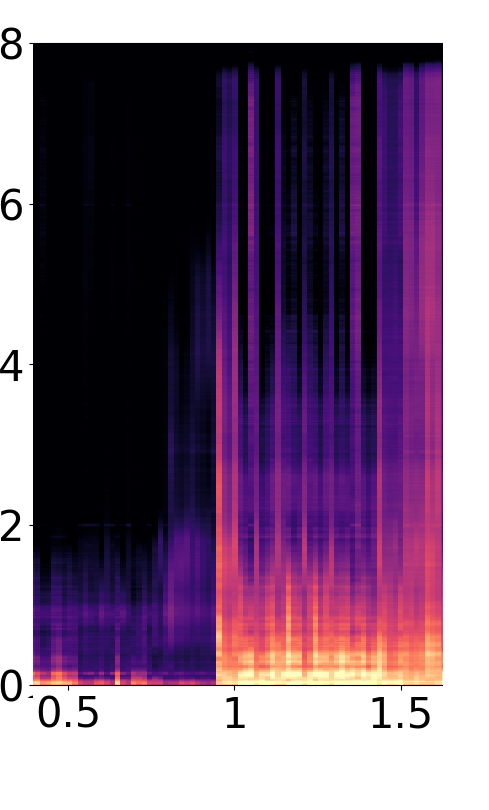} 
}
\hspace{.22cm}
\subfloat[\emph{CVAE} + $\mathcal{L}_\text{adv-VAE}$ \newline w/ $\mathbf{D}_\theta(\mathbf{z_n},y_n)$, $\beta=0$  \label{fig:cvae_ldis}]{
\centering
\includegraphics[width=.342\columnwidth]{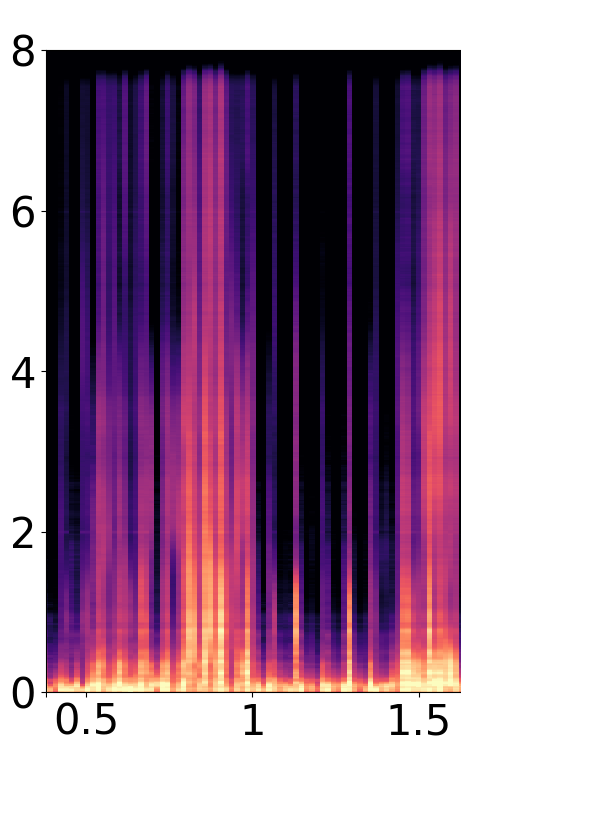}
}
\caption{Reconstructed spectrograms (without MCEM).\label{fig:spectrograms}}
\end{figure}

\begin{table}[t]
\centering
\setlength{\tabcolsep}{3pt}
\begin{tabular}{l | c | c c}
\multicolumn{2}{c}{} & \multicolumn{2}{c}{\textbf{Noise stationarity}} \\
\textbf{Training scheme}&
\textbf{Average} & Nonstationary & Stationary \\
\hline
$\mathcal{L}_\text{adv-VAE}$ & $\mathbf{5.8 \pm 0.1}$ & $\mathbf{5.3 \pm 0.1}$ & $\mathbf{6.8 \pm 0.1}$ \\
\hline
w/ $\mathbf D_{\theta}(\mathbf{z}_n, y_n)$, $\beta=0$ & $5.1 \pm 0.1$ & $4.3 \pm 0.1$ & $6.8 \pm 0.1$ \\
\hline
w/ $\mathbf D_{\theta}(\mathbf{z}_n, y_n)$, $\beta=0$, & \multirow{2}{*}{$5.1 \pm 0.1$} & \multirow{2}{*}{$4.4 \pm 0.1$} & \multirow{2}{*}{$6.4 \pm 0.1$} \\
\phantom{w/ }$\mathcal{L}_\text{adv-enc} = - \mathcal{L}_\text{dis}$ & & \\
\hline
\end{tabular}
\caption{Average SI-SDR (in dB)  w.r.t. different training schemes. \label{tab:training_scheme}}
\label{tab:analysis}
\end{table}

\section{Conclusion}
\label{sec:conclusion}

In this work, we propose to use adversarial training to learn disentanglement between a label describing speech activity and the other latent variables of the VAE. We showed the beneficial effect of learning disentanglement when reconstructing clean speech from noisy speech. In the presence of nonstationary noise, the proposed approach outperforms the standard and conditional VAEs, both trained using the ELBO as loss function, in terms of SI-SDR. Our proposed approach is particularly interesting for audio-visual speech enhancement, where speech activity can be estimated from visual information, which is not affected by the noisy environment.





\bibliographystyle{IEEEtran}
\bibliography{refs21}
%
%
%
%
%
%
%
%
%

\end{sloppy}
\end{document}